\documentclass[
    ,final            
  ]
  {aipproc}

\layoutstyle{6x9}

\begin{document}

\title{Nucleon Structure Functions and Nuclear DIS}


\author{Alexander Molochkov}{
  address={Institut f\"ur Theoretische Physik,
  Justus-Liebig Universit\"at, Giessen, Germany},
  address={Far Eastern National University, Vladivostok, Russia},
  email={alexm@ifit.phys.dvgu.ru},
  thanks={This participation in this conference was sponsored by Alexander
  von Humboldt Foundation}
}

\begin{abstract}
The nucleon structure study in nuclear deep inelastic scattering is considered.
It is shown that nuclear data provide a new source of information about dynamics of
parton distributions in the nucleon. An exmple of the neutron structure function extraction
from the deuteron and proton data is considered. The limit $x\to 1$
of the neutron to proton structure functions ratio is studied.
A link between the deep inelastic scattering off the nucleon
at high $x$ and elastic scattering off nuclei in high $Q^2$ region is discussed.
\end{abstract}

\maketitle

\section{Nucleon structure change in nuclei}

The nuclear data is one of the important sources of information
about nucleon structure.
First of all, it is connected with absence of a free neutron
target for deep inelastic scattering (DIS) experiments, what makes nuclear
data the only source of information about neutron structure functions.
Another important advantage of the nuclear DIS is
access to the kinematic region of large Bjorken $x$ that is unreachable in
DIS off the nucleon. For example, the average nuclear Bjorken $x$
($x_{A}=1/A$)
corresponds to the very large nucleon $x$ ($x_{\rm N}=M_{A}/m_{\rm N}x_{A}\simeq 1$).
However, such study cannot be performed without consistent separation of the nuclear
 and nucleon hard structure. The European Muon Collaboration demonstrated in
 the nuclear DIS experiments
that connection between the nuclear and nucleon hard
structure cannot be trivially explained by the
Fermi-motion of bound nucleons~\cite{aub83}, what was called the EMC-effect.
Study of the $A$-dependence of the effect~\cite{gomez} and nuclear
Drell-Yan scattering~\cite{DY} together
with the theoretical investigations performed within the QMC~\cite{qmc} and
Chiral soliton~\cite{miller} models show that the
EMC-effect cannot be understood without introducing nucleon structure change in
nuclei, which is strongly model dependent.
Recently, the nuclear effects in DIS were studied within a fully covariant
approach~\cite{phlett} based on the Bethe-Salpeter equation. In the paper~\cite{mynucl}
it was found that
the EMC-effect results
 from the Fermi-motion of the bound nucleon in the four-dimensional space, which
leads to the space- and time-smearing of the bound nucleon.
 The time-smearing
 leads to increase of the nucleon localization radius in the four-dimensional
space,
 what corresponds to the explanation of the effect provided by the $Q^2$--rescaling
model~\cite{qresc}.
 In a $3$D projection of the relativistic Fermi-motion the time-smearing
of the
 bound nucleon reduces to the dynamical distortions of the nucleon structure,
which correspond to the nucleon structure change
 obtained within the QMC~\cite{qmc} and Chiral soliton models~\cite{miller}.

 Consistent analytical calculations within this approach in the approximations
 of high $Q^2$ and small binding energy
 give the following expression for the nuclear structure
 function:
\begin{equation}
F_2^{A}(x)=
\int\frac{d^3p}{(2\pi)^3}
\left[\frac{M_A-E_{A-1}-{p_3}}{E_{\rm
N}}F_2^{\rm N}(x_{\rm N})- \frac{\Delta^{\rm N}_{A}}{E_{\rm N}}
x_{\rm N}\frac{d F_2^{\rm N}(x_{\rm N})} {dx_{\rm N}} \right]
\frac{f^{{\rm N}/A}(M_A,{\bf p})}{8M_{A}E_{\rm N}E_{A-1}{\Delta^{\rm
N}_{A}}^2},\label{F2A}
\end{equation}
where $f^{{\rm N}/A}(M_A,{\bf p})$ is a function defined by the solutions of the
Bethe-Salpeter equation, $E_{\rm N}$ is the nucleon on-shell energy,
$\Delta^{\rm N}_A=M_A-E_{\rm N}-E_{A-1}\simeq-T_{kin} +\epsilon$ is the nucleon energy change due to the Fermi motion
($T_{kin} > 0$ -- kinetic
energy of the Fermi motion) and binding
($\epsilon<0$ -- binding energy of the nucleon).

\begin{figure}
\includegraphics[height=.2\textheight]{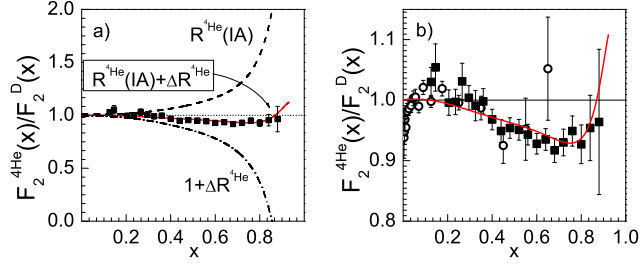}\\
\caption{\label{fig1} Contribution of the impulse approximation (dashed curve
$R^{^4He}(IA)$)
and nucleon structure function derivative (dashed-dot curve $\Delta R^{^4He}$)
to the ratio of the $^4$He and D structure functions.
The experimental
values are shown by the dark squares~\protect\cite{gomez}.}
\end{figure}

Numerical calculations of the contributions
of the different terms in Eq.(\ref{F2A}) to the ratio
$R^{\rm ^4He}=2\sigma^{\rm ^4He}/(A\sigma^{\rm D})=F_2^{\rm ^4He}/F_2^{\rm D}$
are presented at Fig.\ref{fig1} a).
The first term ($R^{\rm ^4He}(IA)$), which results from the Fermi motion in
$3$D-space, is presented by the monotone increasing
dashed curve.
The second term ($\Delta R^{\rm ^4He}$) depicted by the point-dashed curve
results from the time-smearing of the bound nucleon.
Since $M_A<E_{\rm N}+E_{A-1}$ and $F_2^{\rm N}(x_{\rm N})/dx_{\rm N}<0$ this term
puts the ratio $R^{\rm ^4He}$
bellow unity in the middle $x$ region and, therefore, produces the EMC-effect resulted
from cancellation of the two large contributions.
This result is presented by the solid curve, which fits
to the experimental data within the experimental errors with rather good
accuracy
(see Fig.\ref{fig1} b)).
Due to the contribution of $dF^{\rm N}_2(x)/dx$ the expression (\ref{F2A})
establishes a connection between the nuclear data and dynamics of parton
distributions
in the nucleon expressed by the nucleon structure function derivative.

It is worth noting that according to (\ref{F2A}) the functional
behavior of the EMC-ratio is defined by the nucleon structure function properties,
since then the duality observed in $F^{\rm N}_2$
can lead to the dual behavior of the $R^A$ in
the kinematic region of high and low $Q^2$.
This conclusion is consistent with the experimental results obtained recently
in~\cite{arrington}.

\section{Neutron structure function}
Let's consider Eq.(\ref{F2A}) as an integral equation with the
unknown function $F_2^{\rm n}$. The functions $F_2^{\rm p}$
and $F_2^{\rm D}$ are defined by the
existing experimental data. The boundary conditions
 for the equation are fixed with the help of asymptotics
 of the nucleon structure functions
at $x=0$ and $x=1$.
Introducing the neutron structure function $F_2^{\rm n}(x)$ as follows:
\begin{equation}
F_2^{\rm n}(x,Q^2) = r^{\rm n/p}(x)F_2^{\rm p}(x,Q^2);\,\,\,\,\, r^{\rm
n/p}(x)=a_1(1-x)^{\alpha_1}+a_2x^{\alpha_2}+b_1x^{\beta_1}(1-x)^{\beta_2}
(1+c_1x^{\gamma_1})~,
\label{fanzats}\end{equation}
we fix the parameters $a_1$, $a_2$, $\alpha_1$, $b_1$, $\beta_2$ by the boundary conditions.
According to the limit $F_2^{\rm
p}(0)=F_2^{\rm n}(0)$, the parameter $a_1=1$. The parameter $a_2$ is fixed
according to $lim_{x\to 1}F_2^{\rm n}(x)/F_2^{\rm p}(x)$, which is model dependent.
We study
the three limits predicted by the following three different models. The quark model
with SU(6) symmetry gives $r^{\rm n/p}_{x=1} =2/3$.
The elastic limit gives $r^{\rm n/p}_{x=1}=$
$\sigma^{\rm n}_{_{\rm elastic}}/\sigma^{\rm p}_{_{\rm elastic}}\simeq 0.47$.
The minimal value of the ratio has been
predicted by the model with SU(6) symmetry breaking with scalar
diquark dominance --- $r^{\rm n/p}_{x=1} =1/4$.
\begin{figure}
\hspace*{-0.2cm}\mbox{a)\includegraphics[height=.2\textheight]{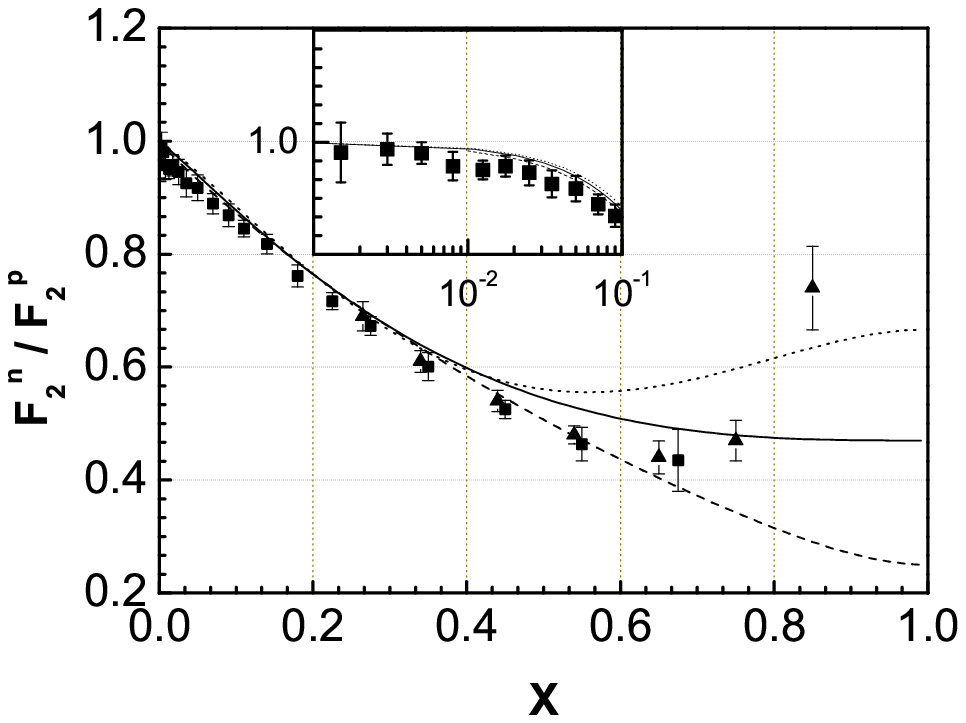}
\hspace*{.1cm}b)\includegraphics[height=.2\textheight]{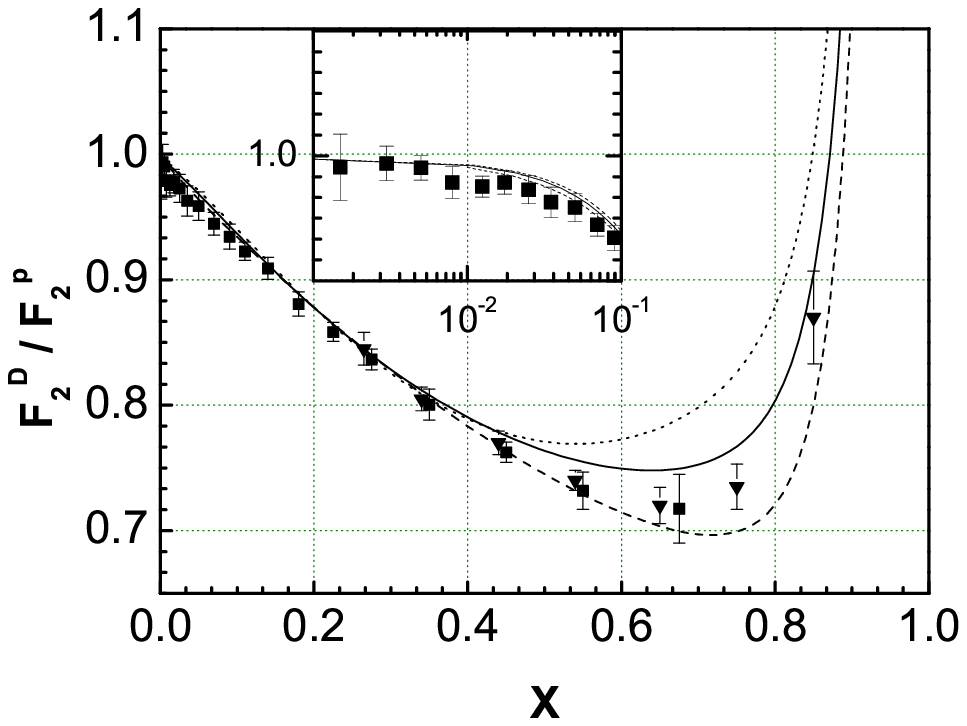}}\\[-1.cm]
\caption{\label{fit}
Comparison of the obtained $F^{\rm n}_2(x)$ with experimental data for
the a) $F^{\rm n}_2/F^{\rm p}_2$ and b) $F^{\rm D}_2/F^{\rm p}_2$ ratios. Experimental data
are taken from SLAC (triangles) and NMC (squares).}
\end{figure}
\begin{figure}
\hspace*{-.5cm}\mbox{\begin{minipage}{6.7cm}
\hspace*{-.3cm}\includegraphics[height=.2\textheight]{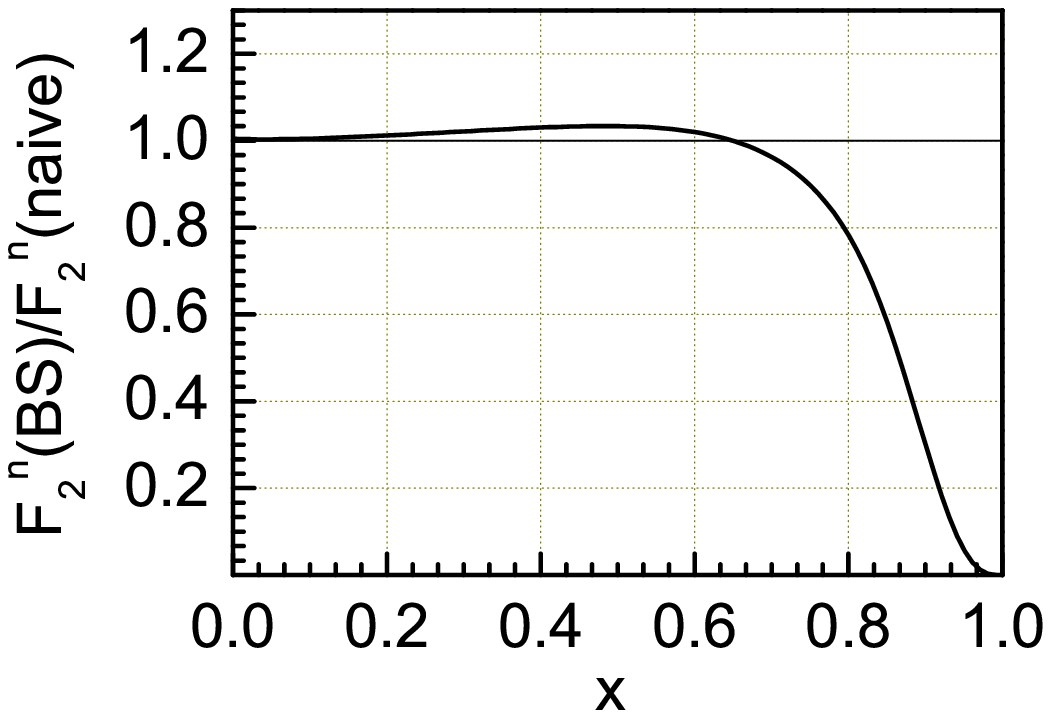}\\
{\footnotesize{\footnotesize\bf FIGURE~3.} The ratio of the $F^{\rm n}_2(x)$
extracted within the present approach to the naive approximation $F^{\rm
n}_2(x)=2F^{\rm D}_2(x)-F^{\rm p}_2(x)$.
}
\end{minipage}\hspace*{.5cm}
\begin{minipage}{6.7cm}
\hspace*{-.3cm}\includegraphics[height=.2\textheight]{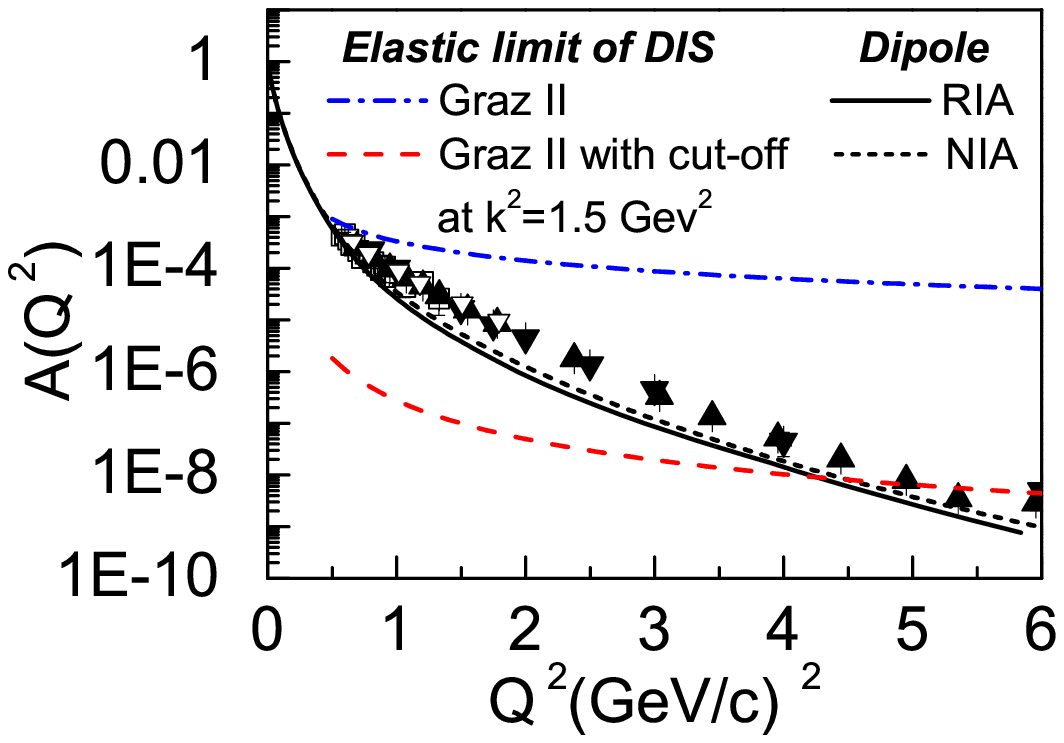}\\
{\footnotesize{\footnotesize\bf FIGURE~4.} Comparison of the elastic limit of
$F^{\rm D}_2(x,Q^2)$ ($x=2$)with the
deuteron structure function $A(Q^2)$.
}
\end{minipage}}
\end{figure}
Assuming
same asymptotic behavior for
$F_2^{\rm p}(x)$  and $F_2^{\rm n}(x)$ at the limit $x\to 1$:
$\lim\limits_{x\to 1}F_2^{\rm p}(x)\simeq C_{\rm
p}(1-x)^3$, $\lim\limits_{x\to 1}F_2^{\rm n}(x)\simeq C_{\rm n}(1-x)^3$;
we get that the derivative of $r^{\rm n/p}$ at $x=1$ is zero.
 It constrains the following parameters of
Eq.~(\ref{fanzats}): $\alpha_1=1$, $\beta_2=1$, $b_1=(\alpha_2
a_2-1)/(1+c_1)$.
The remaining four parameters ($\alpha_2$, $\beta_1$, $c_1$, $\gamma_1$)
are used to fit
Eq.~(\ref{F2A}) to the data for $F^{\rm D}_2(x,Q^2)$ in the range
$10^{-3}<x<0.6$ and $0.5GeV^2<Q^2<40GeV^2$.
The result of the $F^{\rm n}_2$ extraction is presented at Fig.\ref{fit} a).
The experimental data are obtained from the ratio $F^{\rm D}_2/F^{\rm p}_2$
within the naive approximation that neglect all nuclear effects in the deuteron.
Due to the cancellation
of the Fermi-motion and time-smearing
discussed in the previous section the naive approximation works well
up to $x\simeq 0.7$, what explains the good agreement of the obtained here
$F^{\rm n}_2$ with the experimental data. This agreement is illustrated on Fig.3,
which shows that at $x>0.7$ the correct accounting of the
Fermi motion in $4$D space becomes very important.

The obtained nucleon structure function that contains inelastic channels only
allows to study contribution of these channels to the elastic lepton scattering
off the deuteron.
Calculation of the deuteron structure function at the point $x=2$, which
corresponds to the elastic scattering off the deuteron, we get the result
presented at the Fig.4.
Thus, the nucleon resonances, contribution of which is presented by the dashed curve,
can give sizable contribution to the lepton deuteron elastic scattering at high $Q^2$.

\section{Summary}
The nuclear short-range structure can be expressed in terms of the nucleon
structure and
four-dimensional Fermi motion of the nucleons. The time-smearing broadens the
bound nucleon
localization area, what lead to the observation of the nucleon structure change
in nuclei -- EMC-effect. The pattern of the EMC-effect is defined by
cancellation of the two large contributions
that are the $3$D Fermi motion and the time-smearing of the bound nculeons,
which are important at
the average and large Bjorken $x$. Due to the time-smearing the nuclear data
enable direct access to
the information about nucleon structure dynamics expressed by the nucleon
structure function
derivative, $dF^{\rm N}_2(x)/dx$, that is difficult to detect in DIS off the nucleon.
Better accuracy in $F^{\rm D}_2/F^{\rm p}_2$ data at $x\simeq 0.7-0.8$
can give information about the
ratio $F^{\rm n}_2/F^{\rm p}_2$ at $x\to 1$ (see Fig.\ref{fit} b)).
Study of the elastic limit of $F^{\rm D}_2$ shows that nucleon structure
excitations
give sizable contribution to lepton deuteron elastic scattering at $Q^2 >
4GeV^2$.

At the end I would like to thank Alexander von Humboldt Foundation
 for financial support of my participation in this meeting.
 I would like to thank also G.I. Smirnov and U. Mosel
 for useful discussions.
 I am grateful to the organizers of the Workshop for support and
 for warm hospitality during the meeting.

\end{document}